\documentclass{aa}  
\usepackage{graphicx}
\usepackage{txfonts}
\usepackage[colorlinks,citecolor=blue]{hyperref}
\begin{document} 

\title{Dissipation of the striped pulsar wind and non-thermal particle acceleration: 3D PIC simulations}

\author{Beno\^it Cerutti \inst{1}\and Alexander~A. Philippov\inst{2} \and Guillaume Dubus \inst{1}}

\institute{Univ. Grenoble Alpes, CNRS, IPAG, 38000 Grenoble, France\\
           \email{benoit.cerutti@univ-grenoble-alpes.fr}
           \and
           Center for Computational Astrophysics, Flatiron Institute, 162 Fifth Avenue, New York, NY 10010, USA\\
           }

\date{Received \today; accepted \today}

 
\abstract
{The formation of a large-scale current sheet is a generic feature of pulsar magnetospheres. If the magnetic axis is misaligned with the star rotation axis, the current sheet is an oscillatory structure filling an equatorial wedge determined by the inclination angle, known as the striped wind. Relativistic reconnection could lead to significant dissipation of magnetic energy and particle acceleration although the efficiency of this process is debated in this context.}
{In this study, we aim at reconciling global models of pulsar wind dynamics and reconnection in the stripes within the same numerical framework, in order to shed new light on dissipation and particle acceleration in pulsar winds.}
{To this end, we perform large three-dimensional particle-in-cell simulations of a split-monopole magnetosphere, from the stellar surface up to fifty light-cylinder radii away from the pulsar.}
{Plasmoid-dominated reconnection efficiently fragments the current sheet into a dynamical network of interacting flux ropes separated by secondary current sheets which consume the field efficiently at all radii, even past the fast magnetosonic point. Our results suggest there is a universal dissipation radius solely determined by the reconnection rate in the sheet, lying well upstream the termination shock radius in isolated pair producing pulsars. The wind bulk Lorentz factor is much less relativistic than previously thought. In the comoving frame, the wind is composed of hot pairs trapped within flux ropes with a hard broad power-law spectrum, whose maximum energy is limited by the magnetization of the wind at launch.}
{We conclude that the striped wind is most likely fully dissipated when it enters the pulsar wind nebula. The predicted wind particle spectrum after dissipation is reminiscent of the Crab Nebula radio-emitting electrons.}

               
\maketitle


\section{Introduction}

Large-scale current sheets are generic features of planetary and stellar magnetospheres. Their formation can be externally-driven as in the Earth magnetotail shaped by the Solar wind, or internally-driven by the intrinsic magnetic activity of the star or if the magnetosphere is rapidly-rotating like in Jupiter. Perhaps the most extreme example of rotationally-driven current sheets in an astrophysical environment is found in the vicinity of pulsars. The short rotation period of the star ($P\sim1-10^3$ms) combined with strong surface magnetic fields ($B\sim 10^9-10^{12}$G) lead to significant field line winding and opening beyond the light cylinder, a virtual cylindrical surface of radius $R_{\rm LC}=cP/2\pi\sim 50-50,000~$km, beyond which the corotation velocity becomes superluminal. A large-scale current sheet forms outside the light cylinder where both magnetic polarities of the star meet \citep{1971CoASP...3...80M, 1990ApJ...349..538C}. The magnetic and current structures are supported by a plasma of relativistic electron-positron pairs self-generated near the stellar surface via pair production. This plasma flows along open field lines in the form of a radially expanding, relativistic magnetized wind, simply referred as the pulsar wind in the following \citep{1974MNRAS.167....1R, 1984ApJ...283..694K}.  

If the magnetic axis is aligned with the rotation axis, the magnetosphere is axisymmetric and the current sheet is a flat disk lying in the equatorial plane. If the magnetic axis is inclined, the current sheet has the shape of an oscillatory structure of wavelength $2\pi R_{\rm LC}$ confined within an equatorial wedge of half-opening angle set by the magnetic inclination angle \citep{1999A&A...349.1017B}. A cut through this surface at a constant latitude gives rise to a succession of narrow stripes of currents separated by smooth wind regions (Figure~\ref{fig_full}). For this reason, this structure is usually referred to as the ``striped wind'' \citep{1990ApJ...349..538C, 2009ASSL..357..421K}. Away from this region, the wind is smooth and is well described by a rotating monopole-like configuration \citep{1973ApJ...180L.133M}.

One fundamental question refers to the fate of the stripes as the wind propagates outward, and so far this issue has led to contradictory conclusions \citep{1990ApJ...349..538C, 2001ApJ...547..437L, 2003MNRAS.345..153L, 2003ApJ...591..366K, 2017A&A...607A.134C}. It is generally accepted that relativistic reconnection occurs within the current layer (e.g., \citealt{2015SSRv..191..545K}) leading to a transfer of magnetic energy into particle kinetic energy. The main uncertainty lies in the rate of dissipation and its feedback on the global dynamics of the wind. In a collisionless plasma, the current layer thickness is determined by the plasma kinetic scales, i.e., of order the plasma skin-depth and particle Larmor radius scales. In pulsars, this scale is microscopic such that the aspect ratio of the current sheet is very large meaning that the layer will most likely reconnect into the plasmoid-dominated regime, a regime where reconnection is fast \citep{2010PhRvL.105w5002U}. Recent particle-in-cell (PIC) simulations of plane-parallel reconnection have confirmed the efficiency at dissipating the field and at accelerating particles of relativistic reconnection mediated by the plasmoid instability (e.g, \citealt{2001ApJ...562L..63Z, 2012ApJ...754L..33C, 2014ApJ...783L..21S, 2018MNRAS.473.4840W}). In a parallel effort, global PIC simulations of pulsar magnetospheres have shown the major role of reconnection at dissipating a sizeable fraction of the Poynting flux into high-energy particles and pulsed gamma-ray emission \citep{2016MNRAS.457.2401C, 2018ApJ...855...94P, 2018ApJ...857...44K}. These simulations were focused on the inner magnetospheric regions and restricted to a few light-cylinder radii only so that the large-scale evolution of dissipation was not probed.

In this study, we aim at reconciling global models of pulsar wind dynamics, reconnection and particle acceleration in the stripes within the same numerical framework, using large three-dimensional (3D) PIC simulations. The latter are supplemented by a series of two-dimensional (2D) simulations restricted to the equatorial plane to explore the parameter space and the effect of numerical resolution. This work is the logical continuation of our previous effort in this direction \citep{2017A&A...607A.134C}. We begin first by introducing the numerical setup in Sect.~\ref{sect_setup}. Simulation results are presented in Sect.~\ref{sect_results} and discussed in Sect.~\ref{sect_discuss} with an emphasis on dissipation and particle acceleration. Radiative signatures will not be discussed in this paper and will be left to a future study.

\section{Methodology and setup}\label{sect_setup}

We use the relativistic electromagnetic PIC code {\tt ZELTRON} \citep{2013ApJ...770..147C, 2019ascl.soft11012C} in its full 3D spherical coordinates $(r,\theta,\phi)$ version first introduced in \citet{2016MNRAS.457.2401C}. The numerical grid is logarithmically spaced along the $r$-axis. This choice is well-suited for this problem where the plasma density and the field strength present a sharp gradient in the vicinity of the star, and at the same time allows us to probe large physical distances, the key objective in this work. The grid along the $\theta$-direction follows a $\cos\theta$-spacing. This is a natural choice for a 3D spherical grid as it keeps the volume of the cell the same at a given radius, i.e., the grid is refined at the equator but it is coarser at the poles. This choice is also motivated by the fact that the pulsar wind power is concentrated within the equatorial regions (for a monopole it scales as $\propto \sin^2\theta$). The grid is uniformly spaced along the $\phi$-direction.

The full numerical grid is composed of $\left(2016\times 1024\times 512\right)$ cells along the $r$-, $\theta$- and $\phi$-directions respectively. The box extends from the stellar surface $r_{\rm min}=r_{\star}$ up to $r_{\rm max}=100~R_{\rm LC}$ where we fixed $R_{\rm LC}/r_{\star}=3$, $\theta=\left[0.03\pi,0.97\pi\right]$ and $\phi=\left[0,2\pi\right]$. A damping layer absorbs all outgoing electromagnetic waves and particles to mimic an open boundary \citep{2015MNRAS.448..606C}. We apply reflective boundary conditions along the $\theta$-boundaries for the particles and axial symmetry to the fields \citep{1983ITNS...30.4592H}. As for the $\phi$-direction, we apply standard periodic boundary conditions to the fields and the particles. The rotation axis of the star is aligned with the axis of the spherical domain, i.e. $\theta=0$. The magnetic axis is inclined at an angle $\chi$ with respect to the rotation axis and is rotating at the angular velocity of the star $\Omega$. Following \citet{2017A&A...607A.134C}, we chose a split-monopole magnetic configuration \citep{1973ApJ...180L.133M, 1999A&A...349.1017B} which is a good proxy for the asymptotic structure of the pulsar wind which is the main region of interest in this work. At $t=0$, the magnetic field configuration is purely radial,
\begin{eqnarray}
B_{\rm r} = \pm B_{\star}\left(\frac{r_{\star}}{r}\right)^2,
\end{eqnarray}
where $\pm B_{\star}$ is for the northern/southern hemisphere. At any instant $t$, the plane which separates both magnetic polarities at the stellar surface is given by the following condition
\begin{equation}
\sin\theta\sin\chi\cos\left(\Omega t-\phi\right)+\cos\chi\cos\theta=0.
\end{equation}
The solid rotation of field lines is enforced by applying the co-rotation electric field at the stellar surface at every time step. Assuming a perfectly conducting neutron star yields
\begin{equation}
\mathbf{E}=-\frac{\left(\mathbf{\Omega}\times\mathbf{r}\right)\times\mathbf{B}}{c}.
\end{equation}
Fresh plasma exclusively composed of electron-positron pairs is uniformly injected at the stellar surface. The plasma is in co-rotation with the star and has a net radial velocity determined by the $\mathbf{E}\times\mathbf{B}$ drift velocity of the monopole solution. Normalized by the speed of light, the initial particle velocity components are \citep{2017SSRv..207..111C}
\begin{equation}
\mathbf{\beta^{\star}_{\rm r}}=\frac{1}{1+R^2_{\rm LC}/R^2},
\label{eq_vr}
\end{equation}
\begin{equation}
\mathbf{\beta^{\star}_{\rm \theta}}=0,
\end{equation}
\begin{equation}
\mathbf{\beta^{\star}_{\rm \phi}}=\frac{R/R_{\rm LC}}{1+R^2/R^2_{\rm LC}},
\label{eq_vphi}
\end{equation}
where $R=r\sin\theta$ is the cylindrical radius. The injected plasma is neutral and has a multiplicity $\kappa_{\star}=n^{\star}/n^{\star}_{\rm GJ}=10$, where $n^{\star}_{\rm GJ}= \Omega B_{\star}/2\pi e c$ is the fiducial plasma density \citep{1969ApJ...157..869G} and $e$ the elementary electric charge. This prescription is a simple numerical recipe to fill efficiently the magnetosphere with plasma and therefore reach a quasi force-free configuration which is most appropriate to model active pulsars \citep{2015MNRAS.448..606C}. In this work, we are not aiming at modeling pair production which is most likely happening in the inner magnetosphere but focus our numerical resources on the wind region instead, assuming that plenty of pairs are produced along all field lines. On average, one new pair is injected per cell at the neutron star surface every 8 time step. At the end of the simulations when most of the numerical box has been filled with plasma, there are about $\sim 10^{10}$ particles which represents about 10 particles per cell on average.

In addition to the Lorentz force, particles feel the radiation-reaction force due to both curvature and synchrotron radiation (see \citealt{2016MNRAS.457.2401C} for its implementation). The simulations time step is fixed at half the Courant-Friedrichs-Lewy time step, $\Delta t=0.5\Delta t_{\rm CFL}$. One pulsar period is $3.6\times 10^4 \Delta t$. All simulations were evolved for about $10$ rotation periods, i.e., for about $3.6\times 10^5 \Delta t$. The fiducial plasma magnetization at the star surface is set at
\begin{equation}
\sigma_{\star}=\frac{B^2_{\star}}{4\pi n^{\star} m_{\rm e} c^2}=250,
\end{equation}
where $m_{\rm e}$ is the electron rest mass. The shortest plasma time scale, $\omega^{-1}_{\rm pe}$, as well as the shortest radiative cooling time scale, $\omega^{-1}_{c}$, are well resolved in all simulations, with $\omega_{\rm pe}\Delta t \approx 20$, and $\omega_{\rm c}\Delta t \approx 14$. The smallest plasma scale, the local skin depth $d_{\rm e}=c/\omega_{\rm pe}$, is also very well resolved everywhere in the simulation box with $d_{\rm e}/\Delta r\gtrsim 10-30$, where $\Delta r$ is the radial size of a cell. At the light cylinder, the local plasma skin depth is $d_{\rm e}/R_{\rm LC}\approx 3\times 10^{-2}$.

We have performed 3D simulations with three different magnetic obliquity angles: $\chi=30^{\circ}$, $60^{\circ}$ and $85^{\circ}$, all other parameters remaining identical. We ran another 3D run for $\chi=60^{\circ}$ and $r_{\rm max}=50~R_{\rm LC}$ with a constant spacing along the radial direction to evaluate the role of the grid on magnetic dissipation. We have also performed a series of 2D runs limited to the equatorial plane as in \citet{2017A&A...607A.134C} to explore the parameter space and to investigate the role of numerical resolution on the dissipation of the striped wind. To this end, we performed a first series of 6 simulations with $\kappa_{\star}=10$ and $\sigma_{\star}=250$ using a logarithmic or a constant radial grid spacing with different numerical resolution both in $r$ and $\phi$. In a second set, we aim at assessing the role of the wind magnetization $\sigma_{\star}=125$,$~250$,$~750$ while keeping $\kappa_{\star}=10$. In the last set, we explore the role of the plasma multiplicity while keeping the stellar magnetic field the same, with $\kappa_{\star}=0.2$, $0.6$, $2$, $6$, $20$, $60$. Table~\ref{table_sim} draws the full listing of all runs used in this work.

\begin{table}
\caption{\label{table_sim}List of global 3D and 2D PIC simulations of the pulsar striped wind performed in this work.}
\centering
\begin{tabular}{lcccccc}
\hline\hline
Run & Grid cells & $\chi$ & $r_{\rm max}$ & $r$ & $\sigma_{\star}$ & $\kappa_{\star}$\\
\hline
\noalign{\smallskip}
\multicolumn{7}{c}{3D runs: Obliquity \& grid spacing} \\
\noalign{\smallskip}
\hline
Z30 & $2016\times1024\times512$ & $30$ & $100$ & log & $250$ & $10$ \\
Z60 & $2016\times1024\times512$ & $60$ & $100$ & log & $250$ & $10$ \\
Z85 & $2016\times1024\times512$ & $85$ & $100$ & log & $250$ & $10$ \\
Z60c & $2016\times1024\times512$ & $60$ & $50$  & uni & $250$ & $10$ \\
\hline
\noalign{\smallskip}
\multicolumn{7}{c}{2D runs: Resolution \& grid spacing} \\
\noalign{\smallskip}
\hline
Zrl1 & $4032\times2016$ & $90$ & $100$ & log & $250$ & $10$ \\
Zrc1 & $2016\times1008$ & $90$ & $10$ & uni & $250$ & $10$ \\
Zrc2 & $4032\times2016$ & $90$ & $10$ & uni & $250$ & $10$ \\
Zrc3 & $4032\times2016$ & $90$ & $20$ & uni & $250$ & $10$ \\
Zrc4 & $8064\times4032$ & $90$ & $20$ & uni & $250$ & $10$ \\
Zrc5 & $8064\times4032$ & $90$ & $100$ & uni & $250$ & $10$ \\
\hline
\noalign{\smallskip}
\multicolumn{7}{c}{2D runs: Magnetization} \\
\noalign{\smallskip}
\hline
Zs125 & $4096\times2048$ & $90$ & $100$ & log & $125$ & $20$ \\
Zs250 & $4096\times2048$ & $90$ & $100$ & log & $250$ & $20$ \\
Zs750 & $4096\times2048$ & $90$ & $100$ & log & $750$ & $20$ \\
\hline
\noalign{\smallskip}
\multicolumn{7}{c}{2D runs: Multiplicity} \\
\noalign{\smallskip}
\hline
Zk02 & $4096\times2048$ & $90$ & $100$ & log & $250$ & $0.2$ \\
Zk06 & $4096\times2048$ & $90$ & $100$ & log & $250$ & $0.6$ \\
Zk2 & $4096\times2048$ & $90$ & $100$ & log & $250$ & $2$ \\
Zk6 & $4096\times2048$ & $90$ & $100$ & log & $250$ & $6$ \\
Zk20 & $4096\times2048$ & $90$ & $100$ & log & $250$ & $20$ \\
Zk60 & $4096\times2048$ & $90$ & $100$ & log & $250$ & $60$ \\
\hline
\end{tabular}
\tablefoot{2D runs are in the $r\phi$-plane for $\theta=90^{\circ}$. `log' and `uni' stands respectively for logarithmic and uniform grid spacing along the $r$-direction. $r_{\rm max}$ is expressed in units of $R_{\rm LC}$ and the obliquity angle $\chi$ is in degrees.}
\end{table}

\section{Simulation results}\label{sect_results}

\subsection{Plasma structures}\label{sect_plasma}

\begin{figure}
\centering
\includegraphics[width=\hsize]{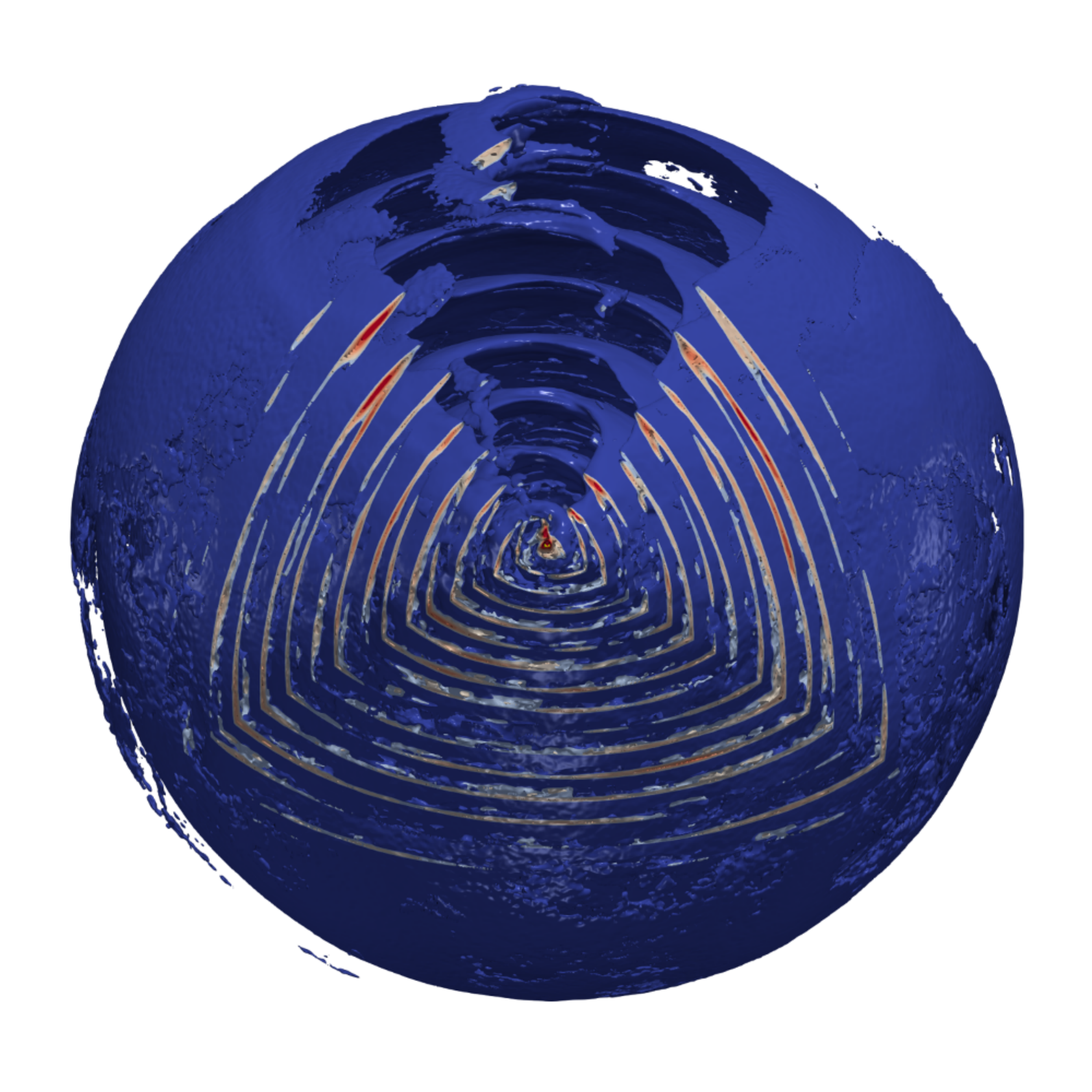}
\caption{3D rendering of plasma density isosurfaces for $\chi=60^{\circ}$ from the star surface (yellow dot at the center) up to $50~R_{\rm LC}$. A spherical wedge has been removed to show the internal structure of the striped wind.}
\label{fig_full}
\end{figure}

\begin{figure*}
\centering
\includegraphics[width=\hsize]{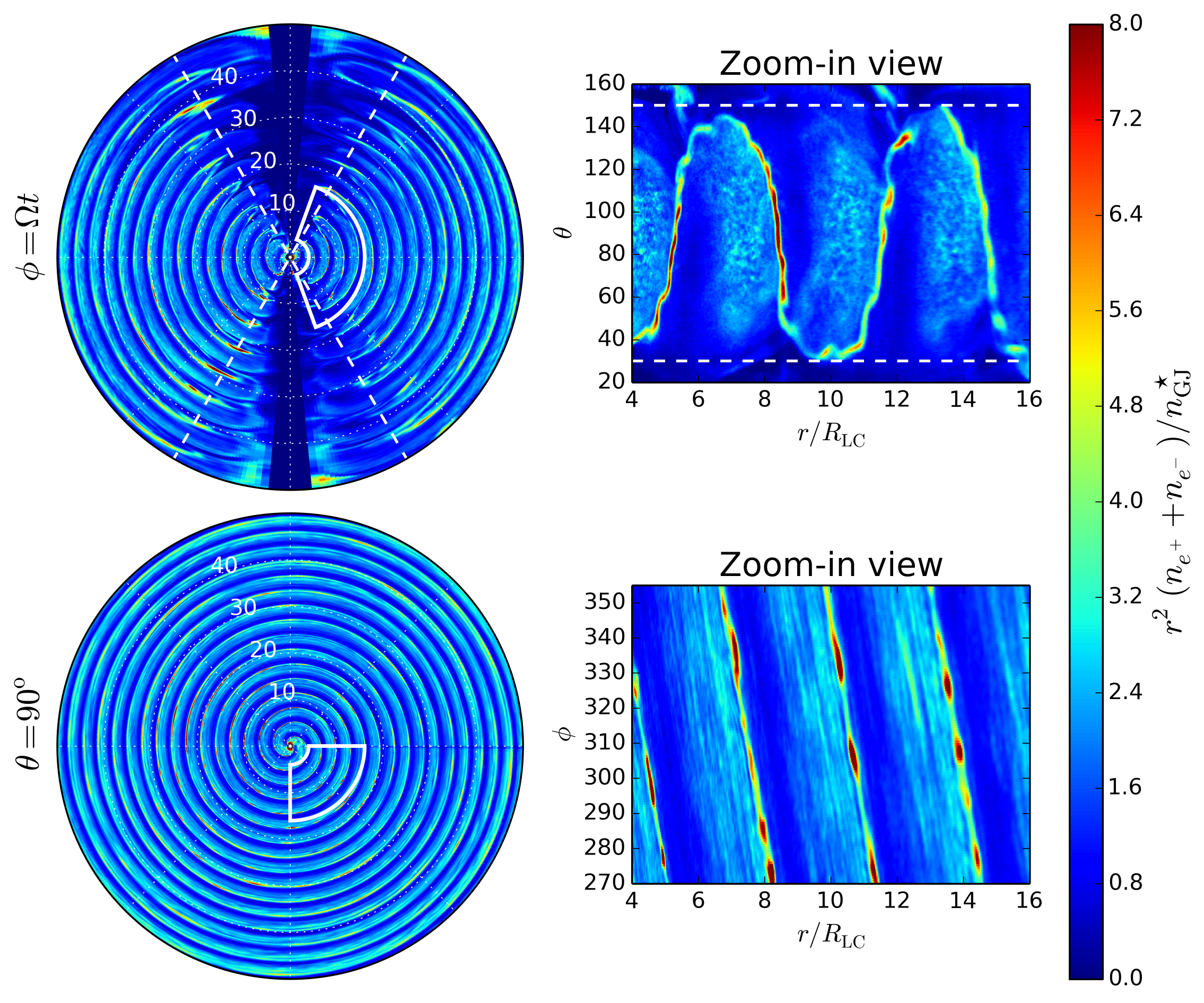}
\caption{2D slices of the plasma density $r^2 (n/n^{\star}_{\rm GJ})$ for $\chi=60^{\circ}$. Top: $r\theta$-plane containing the magnetic axis at the phase $\phi=\Omega t$. Bottom: $r\phi$-plane at the equator ($\theta=90^{\circ}$). The radius is expressed in units of $R_{\rm LC}$. The right panels show zoomed-in views of the regions delimited by the white boxes drawn on the left panels. The striped wind region is contained within $\pi/2-\chi\leq\theta\leq\pi/2+\chi$ (white dashed lines).}
\label{fig_densities}
\end{figure*}

Shortly after the onset of the simulations, the plasma blown from the star establishes a force-free split-monopole-like configuration as it propagates out. It is characterized by a toroidal-dominated magnetic structure whose polarity reverses across the current sheet. The shape of this sheet is consistent with the prediction of \citet{1999A&A...349.1017B}: it is a 3D spherical Archimedean spiral of wavelength $2\pi R_{\rm LC}$ defined within the spherical wedge $\pi/2-\chi\leq\theta\leq\pi/2+\chi$ (Figures~\ref{fig_full}, \ref{fig_densities}). This region, which includes both the sheet and the plasma in between, is the striped wind \citep{1971CoASP...3...80M, 1990ApJ...349..538C, 2009ASSL..357..421K}. Away from the stripes, the wind nearly resembles a single rotating magnetic monopole \citep{1973ApJ...180L.133M}.

The current sheet fragments soon after its formation near the light cylinder. It is unstable to the relativistic tearing instability \citep{1979SvA....23..460Z, 2007PPCF...49.1885P} which mediates fast magnetic reconnection. It results in the formation of a dynamical chain of plasma overdensities trapped in magnetic loops, or plasmoids, separated by secondary current sheets where the field reconnects. These features are clearly visible on the zoomed-in view of the plasma density in both the toroidal and poloidal planes in Figure~\ref{fig_densities}. In full 3D, these structures form a network of interconnected flux ropes reminiscent of plane parallel reconnection simulations \citep{2013ApJ...774...41K, 2014ApJ...782..104C, 2017ApJ...843L..27W}. Secondary flux ropes are being produced within secondary current sheets which then merge with others to form bigger structures. The dynamical nature of reconnection is most pronounced within $\lesssim 10~R_{\rm LC}$ and is quenched by the expansion of the wind at larger radii. Although severely fragmented, the striped wind structure retains its global coherent structure. Hence, the 2D picture drawn in \citet{2017A&A...607A.134C} qualitatively holds in full 3D.

We observe that the sheet is also mildly kink unstable \citep{2005ApJ...618L.111Z, 2014ApJ...782..104C}. It appears as wiggles most visible in the zoomed-in view of the poloidal plane in Figure~\ref{fig_densities}, but these distortions do not lead to the complete disruption of the sheet. The kink is more effective at low magnetic obliquities \citep{2015ApJ...801L..19P, 2016MNRAS.457.2401C} and can lead to a significant latitudinal spreading of the striped wind region, well outside of its natural boundaries. At $\chi=30^{\circ}$, the kink instability may be responsible for a widening of the stripes of about $10^{\circ}$, whereas there are no noticeable deviations for $\chi=60^{\circ}$ and above (Figure~\ref{fig_densities}, top panel). The plasma density is distributed in a highly inhomogeneous and anisotropic manner. The unstriped wind is composed of a low-density uniform plasma, of multiplicity $\kappa=n/n_{\rm GJ}$ of order unity, except close to the axis where numerical plasma fluctuations are artificially enhanced by the spherical grid. In the striped zone, there is a strong plasma density contrast between, in order of increasing density, the wind ($\kappa\sim 2$-$3$), secondary current layers ($\kappa\sim 4$-$5$) and the far more denser flux ropes ($\kappa\sim 10$-$10^3$). The wind zone itself between two stripes is inhomogeneous, in contrast with the unstriped region. There is a clear plasma depletion on the leading edge of the spiral which was already reported in previous studies \citep{2015ApJ...801L..19P, 2017A&A...607A.134C, 2018ApJ...855...94P}. In this sense, reconnection proceeds in a highly asymmetric way.

\subsection{Poynting flux and dissipation}\label{sect_poynting}

\begin{figure*}
\centering
\includegraphics[width=\hsize]{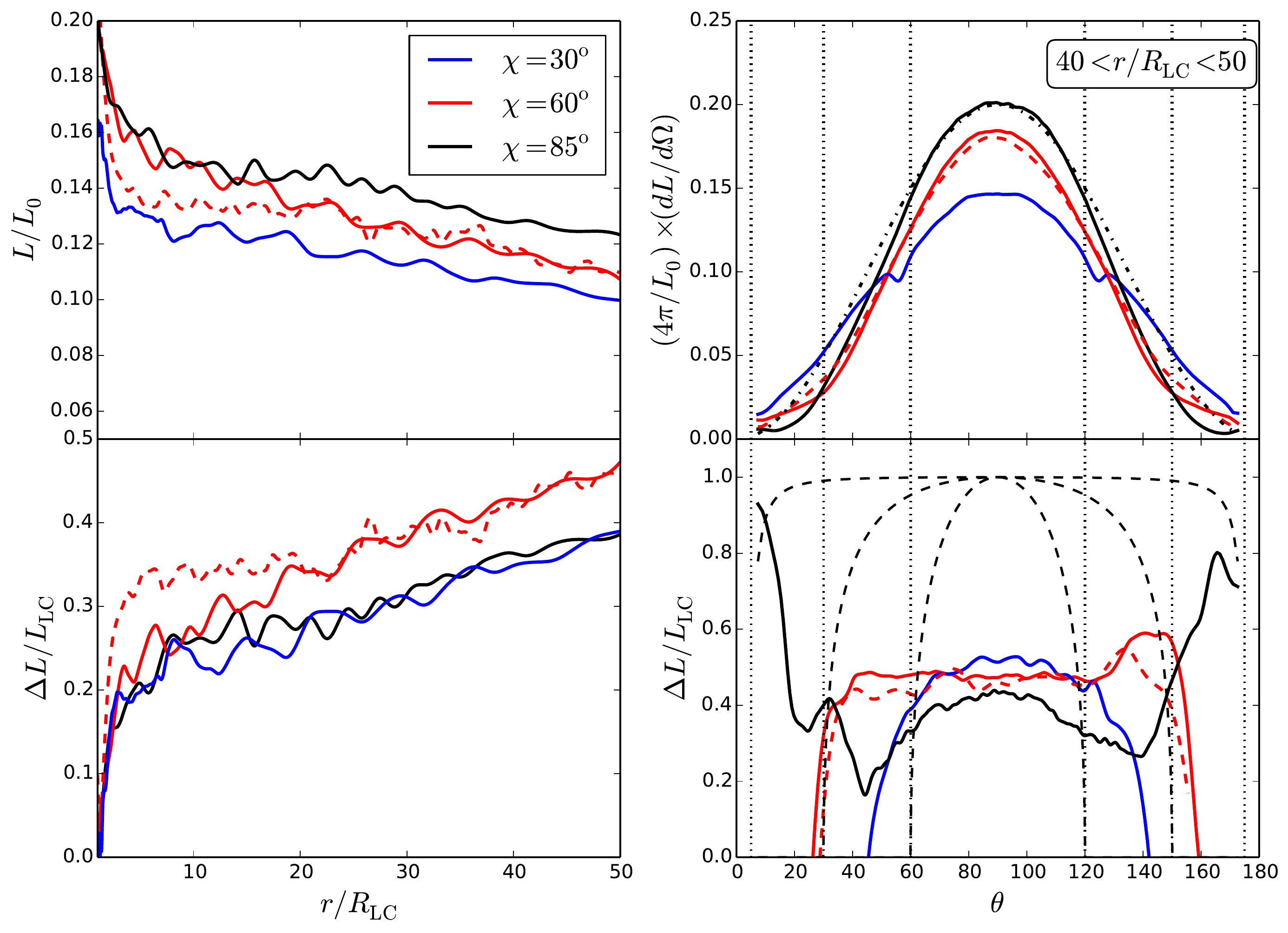}
\caption{Left panels: Radial evolution of the Poynting flux normalized to the expected monopole solution $L_0=2cB^2_{\star}r^4_{\star}/3R^2_{\rm LC}$ (top), and dissipated fraction relative to the light-cylinder value $\Delta L=L(r)-L_{\rm LC}$ normalized by $L_{\rm LC}$ (bottom). Right panels: Latitude dependence of the normalized Poynting flux per unit of solid angle $dL/d\Omega$ (top) and dissipated fraction (bottom) averaged over the radial range $40\leq r/R_{\rm LC}\leq 50$. Vertical dotted lines show the predicted maximal latitudinal extension of the striped wind. The dot-dashed line is a $\sin^2\theta$ profile for comparison with the monopole prediction (top-right panel). The full dissipation model of the striped wind proposed by \citet{2003MNRAS.345..153L, 2013MNRAS.428.2459K} is shown by the black dashed line in the lower-right panel. In all panels, these quantities are shown for all 3D runs performed in this study, including the constant radial grid spacing solution for $\chi=60^{\circ}$ represented by the dashed red line.}
\label{fig_poynting}
\end{figure*}

\begin{figure}
\centering
\includegraphics[width=\hsize]{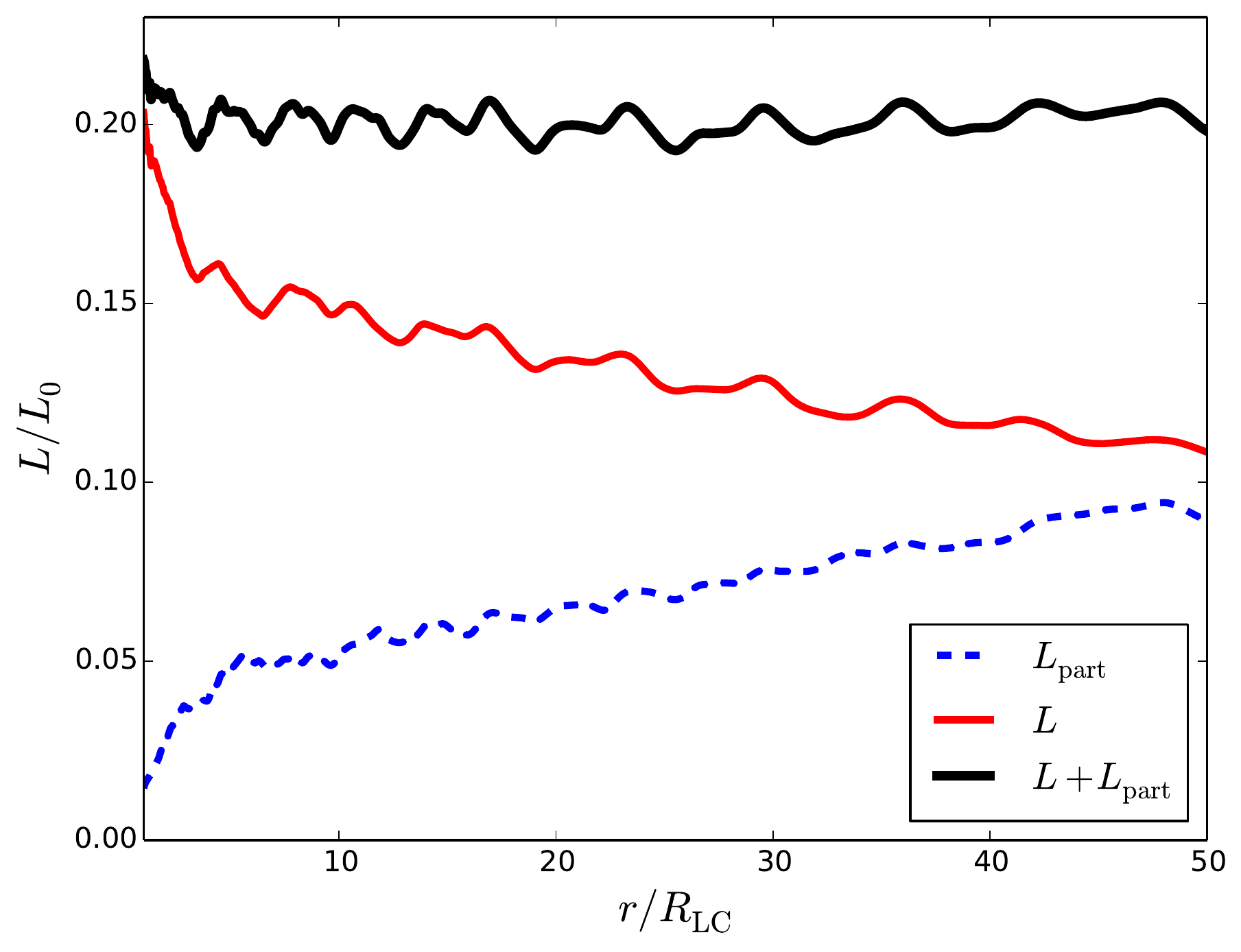}
\caption{Radial evolution of the kinetic energy flux carried by the particles, $L_{\rm part}=\iint n (\gamma-1)m_{\rm e}c^2 v_{\rm r}r^2\sin\theta d\theta d\phi$ (dashed blue line), along with the Poynting flux, $L$ (red solid line), and the total power, $L_{\rm tot}=L+L_{\rm part}$ (thick black solid line) for $\chi=60^{\circ}$. Luminosities are normalized by the monopole spindown power $L_0$.}
\label{fig_lpart}
\end{figure}

\begin{figure}
\centering
\includegraphics[width=\hsize]{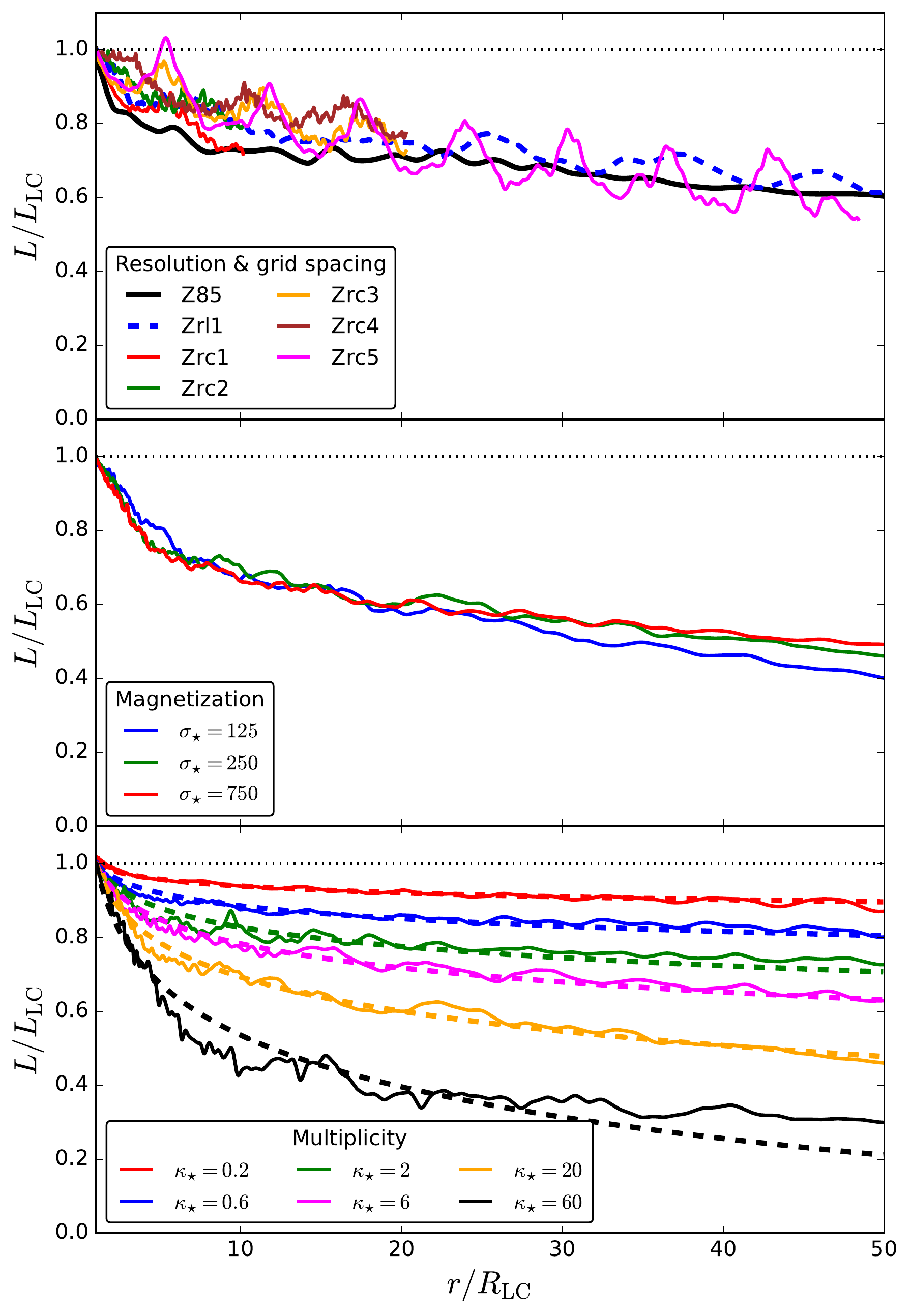}
\caption{Top: Convergence study of dissipation with grid spacing and numerical resolution using 2D simulations (see Table~\ref{table_sim} for grid parameters). The 3D run with $\chi=85^{\circ}$ is reported here for comparison with 2D runs (black solid line). Middle: Dependence of dissipation with the plasma magnetization for $\kappa_{\star}=20$. Bottom: Dependence of dissipation with the plasma multiplicity for $\sigma_{\star}=250$. The dashed lines are best-fit models to the analytical solution proposed in Sect.~\ref{sect_discuss}, Eq.~(\ref{eq_model}). In all panels, the Poynting flux is normalized to its light-cylinder value, $L/L_{\rm LC}$.}
\label{fig_2dpoynting}
\end{figure}

We now turn our attention to the central question of magnetic dissipation. In a dissipationless steady state split-monopole magnetosphere, the outgoing Poynting flux integrated over a spherical radius is conserved in virtue of the Poynting flux theorem. The predicted value is
\begin{equation}
L_{0}=\frac{c}{4\pi}\iint\left(\mathbf{E}\times\mathbf{B}\right)r^2\sin\theta d\theta d\phi=\frac{2cB^2_{\star}r^4_{\star}}{3R^2_{\rm LC}}.
\label{eq_L0}
\end{equation}
Figure~\ref{fig_poynting} shows the radial and latitudinal dependence of the Poynting flux, $L$, for all 3D simulations. The first element to notice is that the numerical values are closer to $L\approx L_0/5$ at the star surface. This discrepancy is explained by the fact that the analytical split-monopole solution used in Eq.~(\ref{eq_L0}) assumes that all field lines cross the light-cylinder and therefore participate to the pulsar spindown. In all runs, a large fraction of initially open field lines reconnects in the equatorial region to form a series of closed field lines nearly co-rotating with the star up to the light cylinder, and thus do not contribute to the outflowing Poynting flux. Only polar field lines remain open so that the magnetospheric structure simulated here qualitatively resembles a force-free dipole inside the light cylinder. Nonetheless, the Poynting flux does not present a strong dependence with magnetic obliquity as expected from the split-monopole solution \citep{1999A&A...349.1017B}. More formally, we can express the spindown power as a function of the flux of open magnetic field lines per hemisphere, $\Psi_{\rm open}$. For an aligned monopole,
\begin{equation}
\Psi_{\rm open}=2\pi r^2 B_{\rm r}\left(1-\cos\theta_{\rm open}\right),
\end{equation}
so that Eq.~(\ref{eq_L0}) should be changed into the more general expression (e.g., \citealt{2016MNRAS.457.3384T})
\begin{equation}
L=\frac{\Omega^2\Psi^2_{\rm open}}{6\pi^2 c}=L_0\left(1-\cos\theta_{\rm open}\right)^2.
\end{equation}
Thus, physically $L/L_0\approx 0.2$ is a measure of the fraction of the solid angle squared filled by open field lines in the inclined split-monopole simulations.

The presence of reconnection in the current sheet leads to significant dissipation which translates into a decay of the radial Poynting flux. Here, dissipation is by no means spurious numerical dissipation but it has a real physical origin: the reservoir of Poynting flux provided by the star is gradually consumed by reconnection which converts magnetic free energy into particle kinetic energy and radiation, such that the total energy in the system is conserved to a good accuracy as shown in Figure~\ref{fig_lpart}. Irregularities in the total power curve reflect the intermittent nature of reconnection. The radial profile of the Poynting flux first drops by about $20\%$ within a few light-cylinder radii after launching, a consistent number with past studies focusing on the magnetosphere and inner wind zone \citep{2012MNRAS.423.1416P, 2014ApJ...785L..33P, 2015MNRAS.448..606C, 2015MNRAS.449.2759B}. Dissipation continues further at a much slower but steady rate beyond $r\gtrsim 10~R_{\rm LC}$ until it reaches about $40$-$50\%$ of the total initial flux at $r=50~R_{\rm LC}$ (Figure~\ref{fig_poynting}). This rate seems independent of the magnetic obliquity. It does not seem to be sensitive on the choice of the grid spacing either: after a brief overshoot at low radii, likely due to the low numerical resolution in comparison with the log-spacing run, the constant $r$-spacing simulation asymptotically converges towards the same amount and rate of dissipation (see the solid and the dashed red lines in Figure~\ref{fig_poynting}). To strengthen this point further, 2D simulations restricted to the equatorial plane with different numerical resolutions, box sizes and grid spacing present the same evolution of Poynting flux and amount of dissipation (top panel in Figure~\ref{fig_2dpoynting}).

Figure~\ref{fig_poynting} also shows the latitudinal dependence of the Poynting flux and of its dissipation. As expected, the Poynting flux is preferentially distributed within the equatorial regions and the $\theta$-profiles closely resemble the split-monopole solution, i.e., $dL/d\Omega\propto \sin^2\theta$. We note that the profiles are slightly sharper than predicted although not as much as reported in \citet{2013MNRAS.435L...1T} where $dL/d\Omega\propto \sin^4\theta$ for an initially dipolar magnetic field configuration. The transition between the striped and the unstriped wind regions is smooth except at $\chi=30^{\circ}$ where small dips are visible. An interesting feature is that the same fraction of power is dissipated at all latitudes within the striped-wind region. This fraction reaches about $40$-$50\%$ depending on the obliquity, and vanishes within the dissipationless unstriped-wind region (see lower-right panel in Figure~\ref{fig_poynting})\footnote{For $\chi=85^{\circ}$ the dissipation rate seems to reach $100\%$ near the $\theta$-boundaries in Figure~\ref{fig_poynting}. This number should be taken with great caution since this is the ratio between two small numbers which are close to numerical fluctuations, and also because it is so close to the spherical axis where the numerics are not free of artefacts.}. Incidentally, the kink-induced widening of the sheet at $\chi=30^{\circ}$ described in Sect.~\ref{sect_plasma} results in a broader angular dissipation rate profile, well outside the boundaries $60^{\circ}\leq\theta\leq 120^{\circ}$.

In Figure~\ref{fig_2dpoynting} (middle and bottom panels), we report on the dependence of dissipation with the plasma magnetization, $\sigma_{\star}$, and multiplicity, $\kappa_{\star}$, based on our large set of 2D simulations. While there is no noticeable dependence with magnetization (at least as long as $\sigma_{\star}\gg 1$), dissipation monotonically increases with increasing plasma multiplicity. Low-multiplicity solutions ($\kappa_{\star}<1$) present 10-20\% dissipation rate at $r/R_{\rm LC}=50$. In contrast, high-multiplicity solution ($\kappa\gg 1$) show up to 70-80\% dissipation without any sign of saturation with radius, indicating that the striped wind structure most likely disappear far before the pulsar wind terminates in isolated systems (Sect.~\ref{sect_discuss}). Putting everything together, simulations reported in this work suggest that reconnection proceeds efficiently and homogeneously within the striped wind, regardless of magnetic inclination, numerical resolution, grid spacing and plasma magnetization, but there is a clear distinction between charge-starved (slow dissipation) and high-multiplicity winds (fast dissipation).

\subsection{Wind kinematics and particle spectra}\label{sect_spectra}

\begin{figure}
\centering
\includegraphics[width=\hsize]{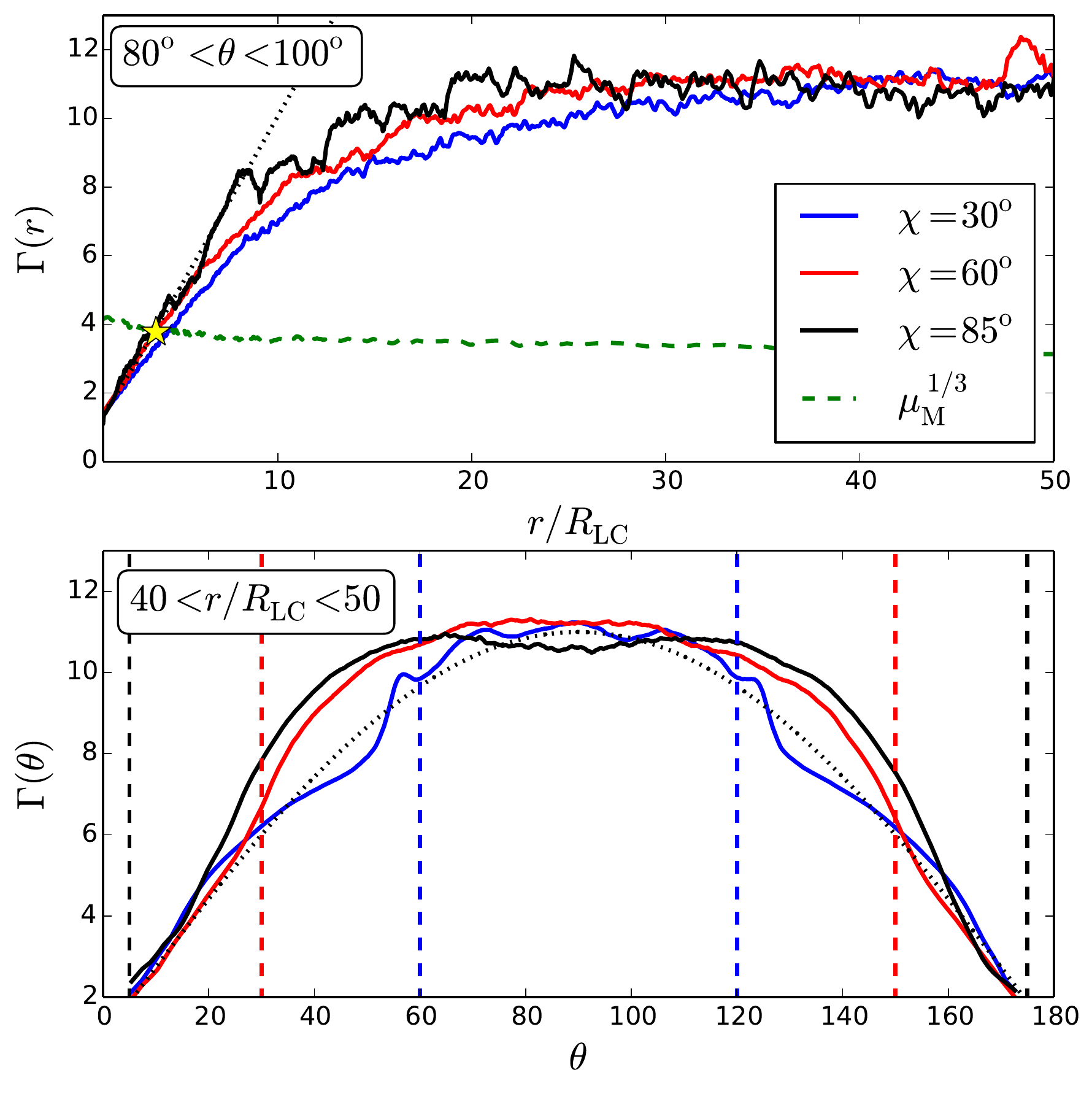}
\caption{Top: $\phi$-averaged plasma bulk Lorentz factor in the striped wind region within $80^{\circ} \leq\theta\leq 100^{\circ}$ as a function of radius, $\Gamma(r)$. The black dotted line is the monopole prediction, $\Gamma=(1+r^2\sin^2\theta/R^2_{\rm LC})^{1/2}$. The green dashed line is the Michel magnetization parameter, $\mu_{\rm M}$ (see Eq.~\ref{eq_mum}), to the power $1/3$. The fast magnetosonic point is located at the yellow star where $\Gamma_{\rm fms}=\mu^{1/3}_{\rm M}$. Bottom: $\theta$-dependence of $\Gamma(\theta)$. Vertical dashed lines delimit the extension of the striped wind region. The black dotted line is a $\sin\theta$ profile as expected from the monopole solution.}
\label{fig_bulk}
\end{figure}

\begin{figure}
\centering
\includegraphics[width=\hsize]{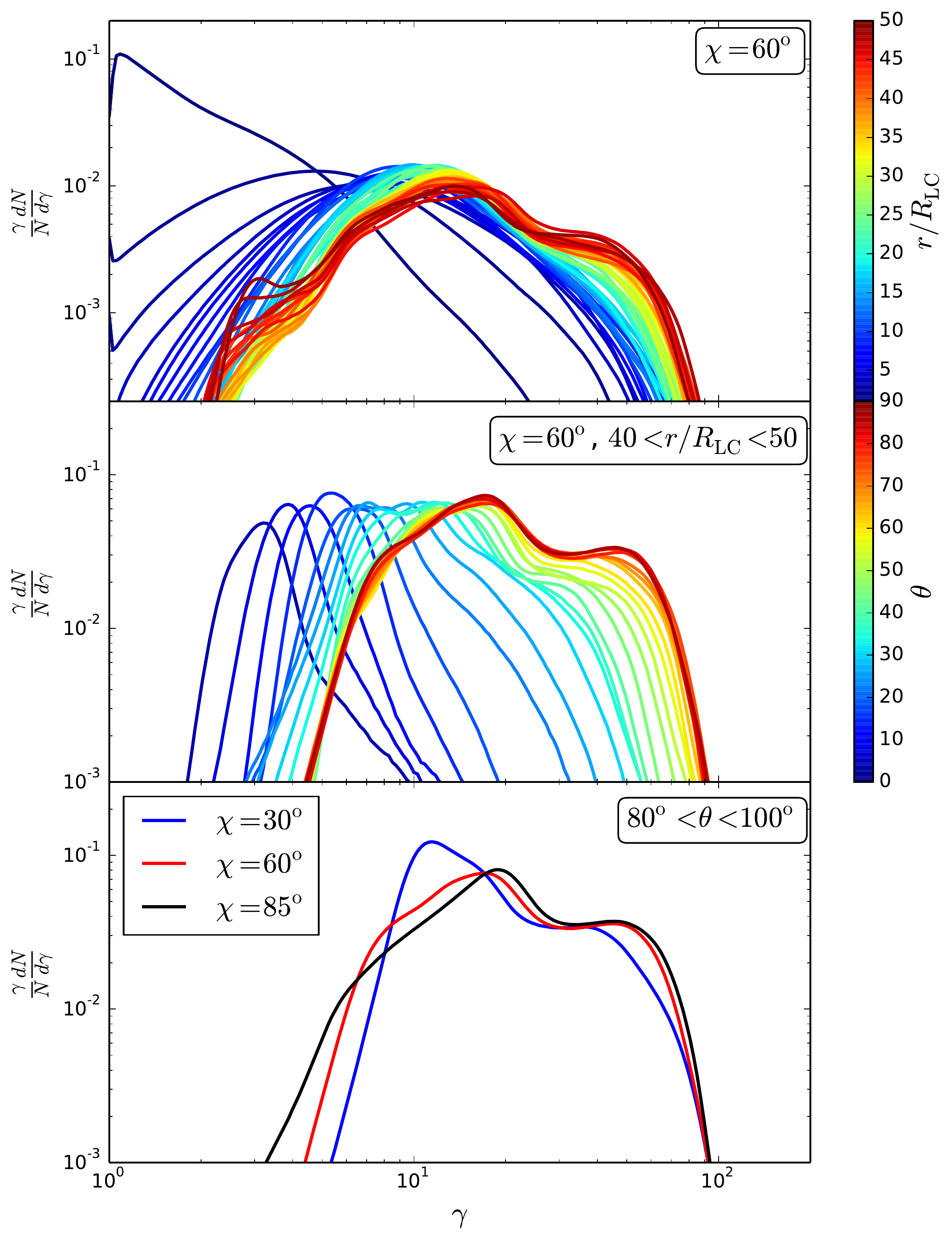}
\caption{Top panel: Normalized particle energy spectrum, $(\gamma/N) dN/d\gamma$, within the spherical shells of radius $r$ and $r+R_{\rm LC}$ as function of radius (color-coded) for $\chi=60^{\circ}$. Middle panel: Latitudinal dependence (color-coded) of the particle energy spectrum averaged between $40\leq r/R_{\rm LC}\leq 50$ for $\chi=60^{\circ}$. Bottom panel: Total particle spectra beyond $r=40~R_{\rm LC}$ and $80^{\circ}\leq\theta\leq 100^{\circ}$ for $\chi=30^{\circ},~60^{\circ}$ and $85^{\circ}$.}
\label{fig_spectra}
\end{figure}

\begin{figure}
\centering
\includegraphics[width=\hsize]{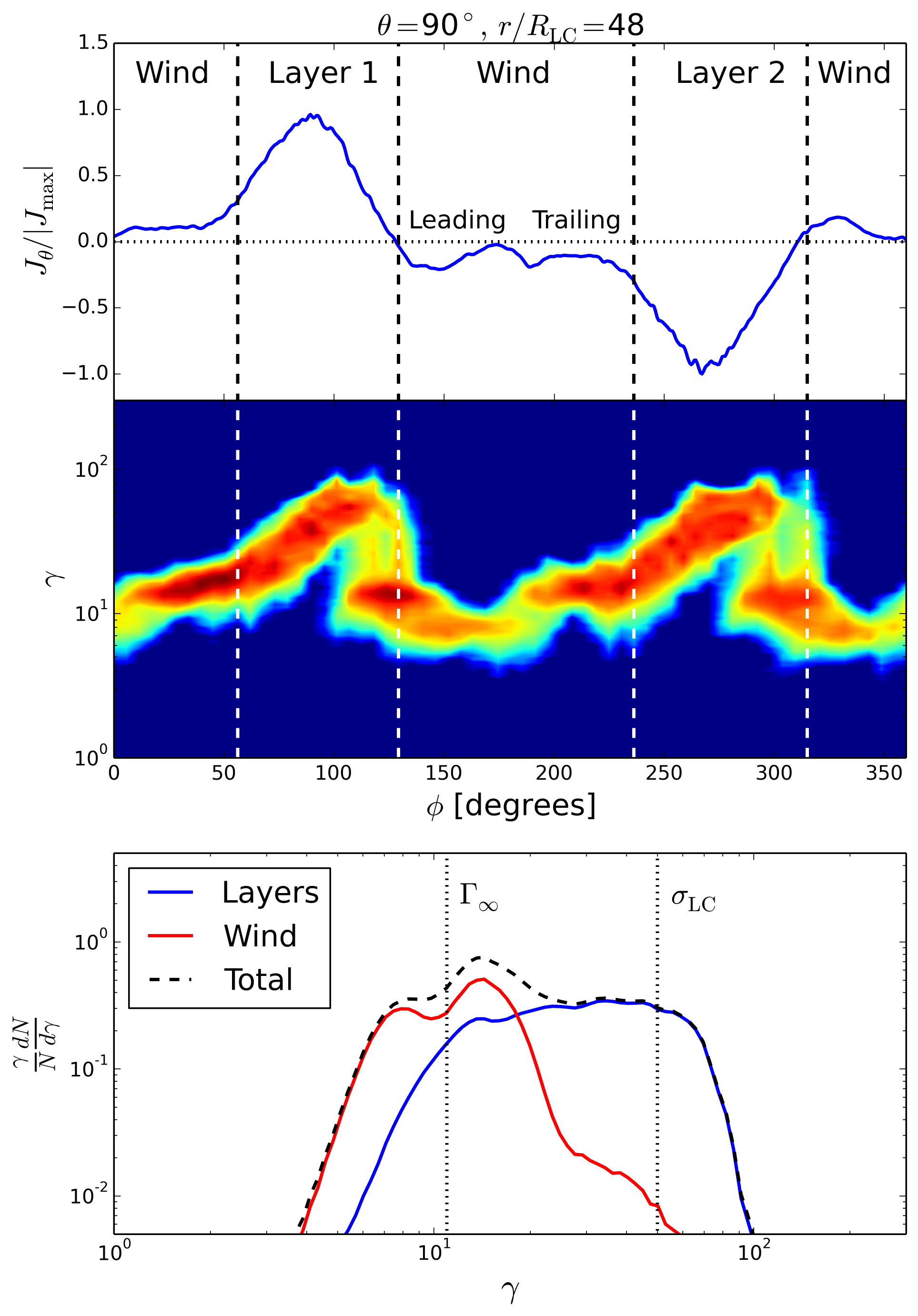}
\caption{Top: $\phi$-profile of the normalized current density flowing along the $\theta$-direction in the equatorial plane ($\theta=90^{\circ}$) for $r=48~R_{\rm LC}$, and $\chi=60^{\circ}$. The locations of the current sheets and wind regions are labelled ``Layer 1'', ``Layer 2'' and ``wind'' respectively. Middle: Corresponding $\phi$-resolved particle energy spectrum. Bottom: Decomposition of the total particle spectrum in the striped region (black dashed line) into a ``wind'' component (solid red line) and a ``Layers'' component (solid blue line). The vertical dotted lines show $\gamma=\Gamma_{\infty}\approx 11$ and $\gamma=\sigma_{\rm LC}\approx 50$.}
\label{fig_map_spectrum}
\end{figure}

In this section, we exploit the particle data to reconstruct the bulk motion of the wind and the particle spectrum in the context of magnetic dissipation and particle acceleration. Figure~\ref{fig_bulk} focuses on the radial and latitudinal dependence of the wind bulk Lorentz factor $\Gamma$. Here, all quantities are averaged over azimuth so that the wind can be referred as a single homogeneous entity. Within $r/R_{\rm LC}\lesssim 4$, the wind accelerates quasi-linearly with cylindrical radius as expected from the monopole prediction, i.e., $\Gamma=(1+r^2\sin^2\theta/R^2_{\rm LC})^{1/2}$ if the plasma follows the $\mathbf{E}\times\mathbf{B}$ drift velocity (see Eqs.~\ref{eq_vr}-\ref{eq_vphi}). Then, the wind quickly reaches the fast magnetosonic point defined by $\Gamma_{\rm fms}=\mu^{1/3}_{\rm M}\approx 4$, where
\begin{equation}
\mu_{\rm M}\equiv \frac{B^2}{4\pi nm_{\rm e} c^2}=\Gamma\sigma,
\label{eq_mum}
\end{equation}
is Michel's magnetization parameter \citep{2009ASSL..357..421K}. With $\sigma\approx 15$ at the fast point, the wind remains highly magnetized. Past this point, the wind acceleration significantly slows down and saturates at $\Gamma_{\infty}\approx 11$ for $r/R_{\rm LC}\gtrsim 30$ within the equatorial regions. This result is not sensitive to the magnetic obliquity angle and confirms theoretical expectations \citep{1994PASJ...46..123T, 1998MNRAS.299..341B}, and our previous findings in 2D although we obtain here significantly larger asymptotic values in full 3D ($\Gamma^{2D}_{\infty}\approx 6$ reported in \citealt{2017A&A...607A.134C}). The latitudinal dependence of the bulk Lorentz factor is consistent with the monopole solution in the unstriped region with $\Gamma(\theta)\propto \sin\theta$. At large radii in the striped region, the profile is much flatter and significantly departs from the ideal solution (bottom panel in Figure~\ref{fig_bulk}). These deviations are most likely due to dissipation and particle acceleration within the current sheets whose effects are not negligible at large radii.

The wind acceleration is an ideal process solely governed by the $\mathbf{E}\times\mathbf{B}$ drift velocity. It should not be confused with non-thermal particle acceleration due to reconnection or other non-ideal dissipative processes which comes on top of this ideal process. Figure~\ref{fig_spectra} shows the radial and latitudinal evolution of the individual particle Lorentz factor spectrum, $(\gamma/N) dN/d\gamma$. The $r$-dependence highlights the formation of a broad non-thermal particle spectrum as magnetic energy is consumed and transferred to the particles via reconnection. The saturated-looking state of the particle spectrum at large radii hides a strong latitude dependence shown in the middle panel of Figure~\ref{fig_spectra}. The spectrum is narrow and peaks at low energies near the rotation axis ($\langle\gamma\rangle\sim 3$). The mean energy shifts to higher energy without any significant spectral broadening in the unstriped zone. This evolution follows the ideal bulk acceleration of the wind which is not accompanied by non-thermal acceleration. In contrast, the particle spectrum dramatically widens inside the striped region where non-thermal particle acceleration pushes particles up to $\gamma\gg\Gamma_{\infty}$. This evolution is not very sensitive to the magnetic obliquity.

The asymptotic spectrum in the striped wind can be decomposed into two parts: a thermal bath peaking at $\gamma\approx \Gamma_{\infty}$ and a hard power-law spectrum $dN/d\gamma\propto\gamma^{-1}$ cutting at $\mu^{\rm LC}_{\rm M}=B^2_{\rm LC}/4\pi n_{\rm LC}m_{\rm e} c^2=\Gamma_{\rm LC}\sigma_{\rm LC}\approx 50$, where $\Gamma_{\rm LC}\approx 1.4$. To uncover the origin of these two components, it is useful to look at the $\phi$-resolved particle spectrum (see Figure~\ref{fig_map_spectrum}). This analysis clearly shows that the low-energy particles are associated with the wind region located far upstream the current layers where the ideal monopole wind acceleration mechanism applies. A closer look reveals that the wind itself is composed of two distinct parts, of slightly different energies, a low-energy component at the leading edge of the spiral sheet corresponding to the low-density region with $\langle\gamma\rangle \approx 8$ and a high-energy component at the trailing edge of the spiral with $\langle\gamma\rangle \approx 15$. There is a sharp spatial segregation about half way in between two current sheets (around $\phi=0^{\circ}$ and $\phi=180^{\circ}$ in Figure~\ref{fig_map_spectrum}) which marks the zone of influence of each reconnecting layers on the plasma in the wind. As for the hard power-law component, it is unambiguously associated with non-thermal particle acceleration within the reconnection layers. The spectral index of $\approx -1$ is consistent with high-$\sigma$ reconnection simulation studies \citep{2001ApJ...562L..63Z, 2014ApJ...783L..21S, 2015ApJ...806..167G, 2016ApJ...816L...8W}. Therefore, to a first order, the asymptotic particle spectrum within the striped wind can be approximatively described as
\begin{equation}
\frac{dN}{d\gamma}\propto\gamma^{-1},~\mu^{1/3}_{\rm M}\lesssim\gamma\lesssim\mu^{\rm LC}_{\rm M}.
\end{equation}
The asymptotic spectrum does not seem to collapse into a narrower distribution at large radii, in contrast to our previous 2D runs suggesting that 3D effects may be important in producing a power-law spectrum.

\section{Interpretation and discussion}\label{sect_discuss}

\subsection{A toy model for dissipation}

\begin{figure}
\centering
\includegraphics[width=\hsize]{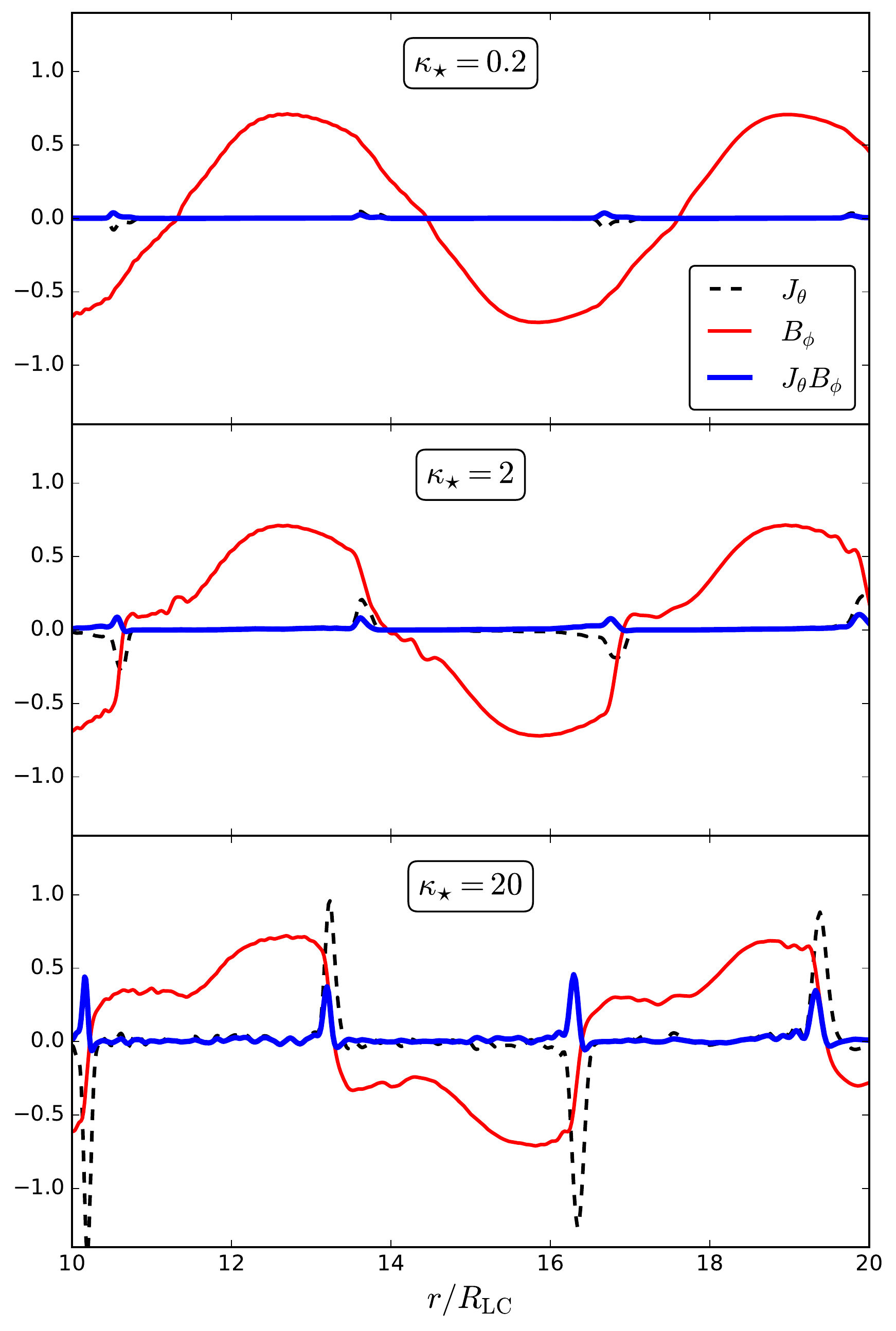}
\caption{Radial profile of $J_{\theta}$ (black dashed line), $B_{\phi}$ (red line) and $J_{\theta}B_{\phi}$ (blue line) for $\kappa_{\star}=0.2$ (top), $2$ (middle) and $20$ (bottom). $J_{\theta}$ is normalized to the Goldreich-Julian current density, $J_{\rm GJ}=(B_{\star}/P)(r_{\star}/r)^2$, and $B_{\phi}$ is normalized by the monopole solution, $B^{\rm Michel}_{\phi}=B_{\star}(R/R_{\rm LC})(r_{\star}/r)^2$.}
\label{fig_edj}
\end{figure}

We propose a simple analytical model for the evolution of magnetic dissipation inspired by simulations in an attempt to extrapolate our results to realistic pulsar winds. In steady state, the radial variations of the Poynting flux is solely governed by Joule dissipation between two spherical shells that we will choose here to be of radius $r=R_{\rm LC}$ and an arbitrary radius $r>R_{\rm LC}$, such that
\begin{equation}
L(r)-L_0=-\int_{R_{\rm LC}}^{r}\int_{0}^{\pi}\int_{0}^{2\pi} \mathbf{J}\cdot\mathbf{E}~r^2\sin\theta dr d\theta d\phi.
\label{eq_joule}
\end{equation}
To a very good accuracy, we have
\begin{equation}
\mathbf{J}\cdot\mathbf{E}\approx J_{\theta}E_{\theta}\approx J_{\theta}B_{\phi}.
\end{equation}
To make further progress, we need to have a closer look at the distribution of currents and fields in the vicinity of current sheets where Joule dissipation is localized. Figure~\ref{fig_edj} shows the radial profile of $B_{\phi}$, $J_{\theta}$ and the product of the two for $\kappa_{\star}=0.2,$ $2$, and $20$ at a pulsar phase $\phi$ chosen such that it crosses only currents sheets and avoids magnetic islands. The spatial distribution of $\mathbf{J}\cdot\mathbf{E}$ within islands is dipolar such that there is no net contribution integrated over their volume, in contrast with secondary current sheets. The magnitude of the current carried by the particles depends on the plasma multiplicity and is localized in the form of thin sheets as expected. At low multiplicities ($\kappa_{\star}<1$), the conduction current is small, too small in fact to explain the magnetic reversal and it is not localized at the magnetic nulls. In this regime, the current is mostly carried by the displacement current ($\partial E/\partial t$, only possible for an oblique rotator), which is a logical consequence of charge starvation in the wind. With increasing multiplicities, the magnetic profile changes from a nearly sinusoidal shape at low-$\kappa_{\star}$, to an asymmetric square shape at high-$\kappa_{\star}$ with sharp gradients where the field reverses and where a strong conduction current is localized. Thus, the electric field in Joule's term may be understood as the reconnection electric field, $E_{\rm rec}$, which represents a fraction of the upstream magnetic field, $B^{\rm up}_{\phi}$, such that
\begin{equation}
E_{\rm rec}=\beta_{\rm rec}B^{\rm up}_{\phi},
\end{equation}
where $\beta_{\rm rec}$ is the dimensionless reconnection rate \citep{1996A&A...311..172L, 2014ApJ...780....3U}.

\begin{figure}
\centering
\includegraphics[width=\hsize]{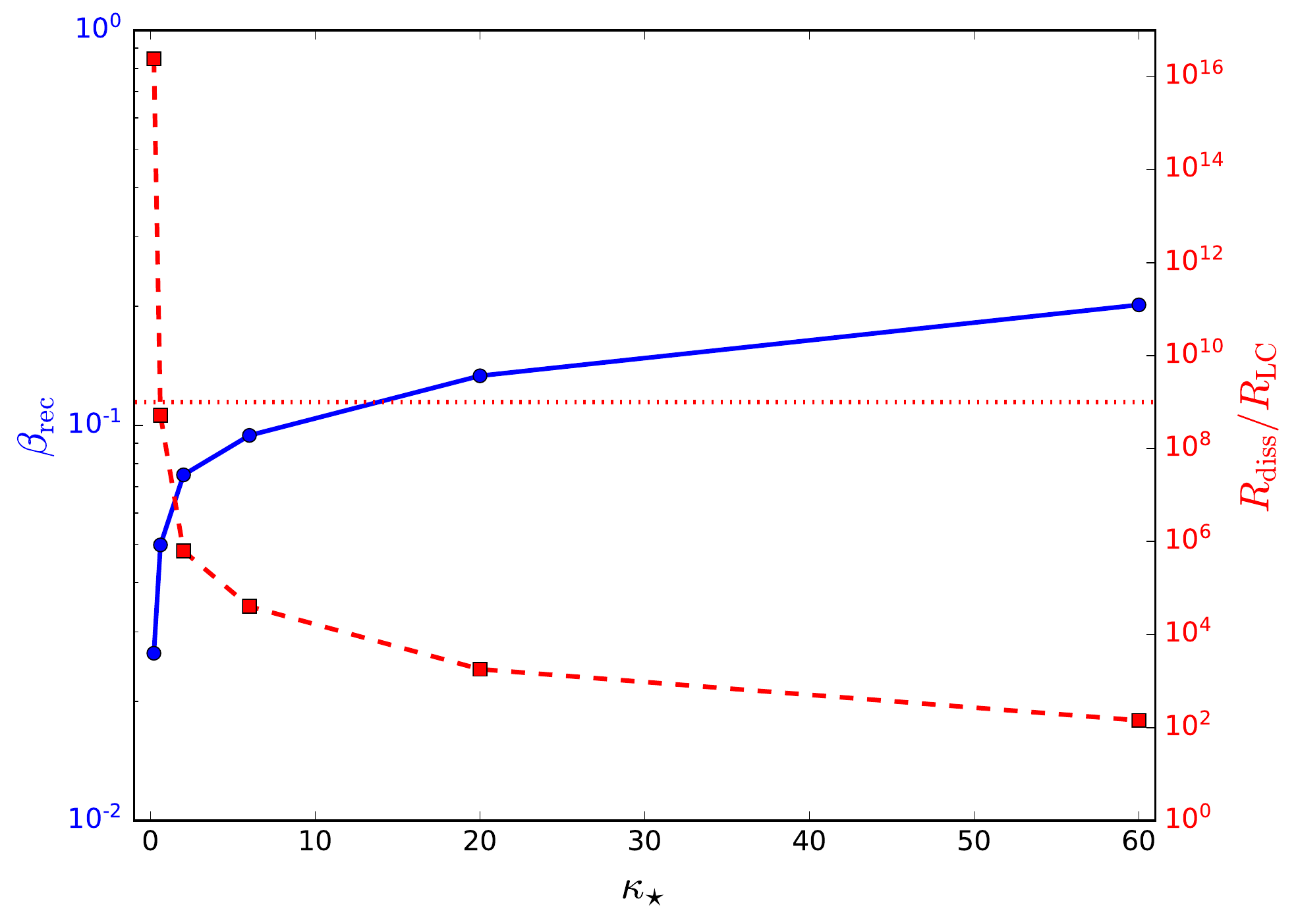}
\caption{Evolution of the dissipation rate $\beta_{\rm rec}$ (blue line, dots and axis) and the extrapolated full dissipation radius of the stripes $R_{\rm diss}/R_{\rm LC}=\exp(\beta^{-1}_{\rm rec})$ (red line, dots and axis) with the typical plasma multiplicity measured at secondary current sheets, $\kappa_{\rm X}$. The estimated radius of the Crab pulsar wind termination shock is shown by the horizontal dotted line ($R_{\rm TS}/R_{\rm LC}\sim 10^9$) for comparison.}
\label{fig_alpha}
\end{figure}

Another important observation we can make from Figure~\ref{fig_edj} is that the magnetic field strength which contributes to dissipation is always on the trailing edge of the spiral because of the asymmetric nature of reconnection in the striped wind, the most pronounced effect being visible at high multiplicities. This conclusion is also compatible with the asymmetry in the phase-resolved particle spectrum reported in Sect.~\ref{sect_spectra}. We find that the upstream magnetic field strength at the trailing edge of the sheet, $B^{\rm up}_{\phi}$, is of order the ideal split monopole field at all radii, even though the striped-averaged field strength decreases with radius due to dissipation, i.e.,
\begin{equation}
B^{\rm up}_{\phi}\approx B^{\rm Michel}_{\phi}= \mp \frac{R}{R_{\rm LC}}B_{\star}\left(\frac{r_{\star}}{r}\right)^2,
\end{equation}
such that the sheet is always fed with fresh, unreconnected magnetic field. The electric current is given by Amp\`ere's law. Assuming perfect symmetry on both sides of the current layer for the sake of simplicity, we have (e.g., \citealt{2017A&A...607A.134C})
\begin{equation}
\frac{4\pi}{c}J_{\theta}\delta=2B^{\rm up}_{\phi},
\end{equation}
where $\delta$ is the layer thickness. Putting everything together, Joule's term within a single current sheet can be approximately estimated as
\begin{equation}
\mathbf{J}\cdot\mathbf{E}\approx \frac{c\beta_{\rm rec}(B^{\rm up}_{\phi})^2}{2\pi\delta}=\frac{3\beta_{\rm rec}L_0}{4\pi\delta}\frac{\sin^2\theta}{r^2}.
\end{equation}
The integral over $\phi$ at constant radius and latitude vanishes everywhere, except at two phases where the layers are located. The angular width of each layer can be estimated as $\Delta\phi\sim\delta/r$, such that Eq.~(\ref{eq_joule}) becomes
\begin{equation}
L(r)-L_0\approx-\frac{3\beta_{\rm rec}L_0}{2\pi}\int_{R_{\rm LC}}^r\int_0^{\pi}\frac{dr}{r}\sin^3\theta d\theta,
\end{equation}
i.e., dissipation is independent of the width of the current layer. The integrals over $\theta$ and $r$ lead to the final result,
\begin{equation}
\frac{L(r)}{L_0}=1-\beta_{\rm rec}\ln\left(\frac{r}{R_{\rm LC}}\right),
\label{eq_model}
\end{equation}
where we absorbed the constant $2/\pi$ of order unity into the reconnection rate, i.e., $\beta_{\rm rec}\leftarrow 2\beta_{\rm rec}/\pi$. Applying this model to simulation data provides a good description of the Poynting flux decay and a direct measure of the rate of dissipation. Figure~\ref{fig_alpha} shows the evolution of the dissipation rate with the plasma multiplicity. It is determined from the best-fit model to the dissipation curves in Figure~\ref{fig_2dpoynting} (bottom panel), assuming the evolution given in Eq.~(\ref{eq_model}). As expected, at low multiplicities dissipation is low because the current is mostly carried away by the displacement current. The contribution from the conduction current increases with plasma supply until it reaches an approximate saturation at high-multiplicities where $\beta_{\rm rec}\sim 0.1$ -- $0.2$. One should keep in mind that the rate measured from the simulations includes additionnal physical effects neglected in the above toy model which are of order unity, such as asymmetries, deviations from the split monopole solutions, or the filling factor of plasmoids which do not contribute to dissipation. These effects may account for the slow increase of the dissipation rate at high multiplicities.

Another effect not captured by this simple model is dissipation due to the formation of vacuum gaps in the low-multiplicity solutions, as reported in \citet{2015MNRAS.448..606C} for an aligned rotator. In the 2D equatorial plane simulations used here, $\boldsymbol{\Omega}\cdot\mathbf{B}=0$ and therefore no gap can form in this special configuration. We would expect an additional source of dissipation at low, but finite multiplicities in a full 3D setup, which we did not explore in this work. In this sense, the model gives a lower limit of dissipation in this regime. This being said, the real solution must smoothly connect to the fully dissipationless vacuum solution \citep{1955AnAp...18....1D}, meaning that dissipation should necessarily cease as we asymptotically approach the vacuum regime, as reported here.

\subsection{Implications}

If the model presented above is a fair description of real astrophysical pulsar winds, it has important astrophysical implications. The first observation to make is that dissipation depends only on the reconnection rate. Studies of local plane-parallel reconnection consistently find a high reconnection rate, $\beta_{\rm rec}\sim 0.1$ -- $0.2$ (see, e.g., \citealt{2015SSRv..191..545K} and references therein), similar to what is reported here. This rate does not seem to depend on anything as long as the layer is thin, meaning that it is of the order the plasma kinetic scales, $\delta\sim d_{\rm e}$. In particular, it should not depend on the plasma multiplicity as long as $\kappa\gg 1$ as reported in Figure~\ref{fig_alpha}, and thus, the radius at which the striped wind has fully dissipated should be similar for all pulsars producing pairs, i.e., like gamma-ray pulsars. In this sense, our results suggest that there is a universal dissipation radius for all pulsar winds loaded with pairs. Using Eq.~(\ref{eq_model}) and assuming a naive extrapolation of the model to any radius, we predict a complete decay of the Poynting flux at a radius
\begin{equation}
R_{\rm diss}=R_{\rm LC}\exp\left(\beta^{-1}_{\rm rec}\right)\sim 10^2 - 10^4 R_{\rm LC},
\label{eq_rdiss}
\end{equation}
for $\beta_{\rm rec}=0.1$ -- $0.2$. Although this radius is quite sensitive to the exact value of the reconnection rate, simulations show that $\beta_{\rm rec}$ is high meaning that the striped wind will most certainly decay entirely far before reaching its termination \citep{1990ApJ...349..538C, 2017A&A...607A.134C}, unless it is truncated at short distances by an accretion disk or a companion star wind.

At this stage, it is important to emphasize that the expansion of the current layer plays no role at dissipating the field, as long as the layer thickness remains small compared with the stripe half-wavelength, $\delta/\pi R_{\rm LC}\ll 1$ \citep{2001ApJ...547..437L, 2003ApJ...591..366K, 2017ApJ...847...57Z}, which is the case in all of the simulations reported here. In contrast to Eq.~(\ref{eq_rdiss}), this condition explicitly depends on the plasma multiplicity: a denser sheet is also a thinner one, thus the condition for which two consecutive sheets would overlap is pushed further away as the multiplicity increases \citep{2017A&A...607A.134C}. Therefore, $R_{\rm diss}$ should be seen as an upper limit for the radius of full dissipation. It is also worth noting that we do not find any evidence for a significant bulk acceleration of the wind due to magnetic dissipation as anticipated by \citet{2001ApJ...547..437L} and \citet{2003ApJ...591..366K}. Here, reconnection proceeds at a similar rate even past the fast point where the bulk Lorentz factor remains constant (Figure~\ref{fig_bulk}). As discussed further below, dissipation does not perform work on the wind as a bulk but rather benefits to disorganized energetic pairs trapped within plasmoids.

The wind kinematics and the shape of the particle spectrum reported here also have important astrophysical consequences. The commonly accepted picture is that the wind is composed of cold, nearly monoenergetic pairs travelling at ultra-relativistic velocities such that $\Gamma\sim 10^4-10^6$ \citep{1974MNRAS.167....1R, 1978MNRAS.185..297W, 1984ApJ...283..694K}. While most of the predicted wind properties are recovered here, we find that the bulk Lorentz factor may not be as relativistic as previously thought. A good proxy is given by the wind Lorentz factor at the fast magnetosonic point, $\Gamma_{\infty}\sim \mu^{1/3}_{\rm M}$. The magnetization at the light cylinder is poorly constrained mostly because of the uncertainty on the plasma multiplicity, but it could of order
\begin{equation}
\mu^{\rm LC}_{\rm M}=\frac{B_{\rm LC}^2}{4\pi \kappa n_{\rm GJ}m_{\rm e} c^2}=\frac{ePB_{\rm LC}}{4\pi m_{\rm e} c\kappa}=1.4\times10^{6}P_{100}B_{5}\kappa^{-1}_4,
\end{equation}
where $P_{100}=P/100~$ms is the pulsar period, $B_5=B_{\rm LC}/10^5$G and $\kappa_4=\kappa/10^4$, for a typical young gamma-ray pulsar. Hence, we expect $\Gamma_{\infty}\sim\mu^{1/3}_{\rm M}\sim 100$ at most, i.e., more similar to what is usually inferred in gamma-ray burst jets\footnote{The cascade developing at the polar caps, which is not captured here, may give an additional bulk motion to the plasma flow injected near the stellar surface, and thus may increase this upper estimate.}. This gives a good estimate of the bulk particle Lorentz factor throughout the pulsar wind. It is also a fair estimate of the individual particle Lorentz factor located within ideal regions, i.e., the unstriped polar regions and the inter-stripe medium where the particle spectrum remains narrow and cold. Within the current layers, relativistic reconnection leads to the formation of a broad power law of index $p\sim -1$, with a low-energy cut off at $\gamma\approx\Gamma_{\infty} \sim 10^2$, and a high-energy cut off at $\gamma\approx \mu^{\rm LC}_{\rm M}\sim 10^6$.

The fundamental difference with the standard picture is that the pairs accelerated in the sheet are not cold with $\gamma=\Gamma$, but instead the particles remain hot and trapped within the flux ropes in the wind frame. Thus, we expect the pulsar wind to inject ultra-relativistic pairs with a hard spectrum into the nebula and not just at a single energy scale. Interestingly, applying this model to the Crab Nebula with the typical magnetization scale quoted above $\mu_{\rm M}\sim 10^6$ could provide a natural explanation for the mysterious radio-electron component responsible for the hard, low-energy emission from the nebula. In the spectral modeling by \citet{2010A&A...523A...2M}, this population is well fitted by a single power law of index $-1.6$ cutting off at $\gamma_{\rm min}\sim 20$ and $\gamma_{\rm max}\approx 2\times 10^5$ which fits well within our results. Although our spectrum is slightly too hard, a larger and therefore more realistic separation of scales between the layer thickness and the light-cylinder scale could lead to a significant spectral steepening \citep{2018MNRAS.481.5687P}. Pairs injected into the shock could be further accelerated into the nebula by some other mechanisms to form the softer X-ray to gamma-ray electron component which would naturally result in a smooth transition between both components.

\section{Summary}

We have performed large 3D PIC simulations of pulsar winds with a focus on magnetic dissipation and particle acceleration within the striped region. The global structure of the striped wind is consistent with the split-monopole prediction \citep{1973ApJ...180L.133M, 1999A&A...349.1017B}. The current sheet is prone to the plasmoid instability shortly after its launching at the light cylinder. This instability leads to an efficient fragmentation of the sheet into a network of interconnected flux ropes separated by secondary thin current layers where the field reconnects, a structure reminiscent of 3D plane-parallel reconnection studies. This chain is highly dynamical, flux ropes form and merge to form bigger structures, which may result in bright short bursts of radio emission \citep{2019ApJ...876L...6P, 2019MNRAS.483.1731L} aligned with the incoherent gamma-ray pulsed emission from the sheet \citep{2016MNRAS.457.2401C}. The sheet is also kink unstable which leads to a significant widening of the striped wind region at low magnetic obliquity. Reconnection in this environment proceeds in a highly asymmetric way, in a qualitatively similar manner as in the dayside of the Earth magnetopause. More efforts are needed to better understand asymmetric reconnection in the relativistic regime using local studies.

Relativistic reconnection gradually consumes the oscillatory component of the toroidal magnetic field. This feature is robust against numerical resolution, grid spacing and plasma magnetization. The Poynting flux monotonically decreases with the distance to the light cylinder, reaching up to about $40\%$ dissipation at the outer parts of the box, i.e., $r=50~R_{\rm LC}$. The dissipation rate weakly depends on latitude within the striped region and on the magnetic obliquity angle. Based on a large set of 2D simulations restricted to the equatorial plane, we can establish that the fate of the striped wind is not sealed by the sheet width as previously thought, but it is rather controlled by the dimensionless reconnection rate $\beta_{\rm rec}\approx 0.2$ which lies well within reported values in plane parallel reconnection studies where $\beta_{\rm rec}\sim 0.1-0.2$. This rate is known to weakly depend on the system size and plasma magnetization in the ultra-relativistic regime ($\sigma\gg1$, \citealt{2018MNRAS.473.4840W}), meaning here that dissipation should not depend on other parameters such as the plasma multiplicity as long as there is a large supply of pairs ($\kappa\gg 1$). Therefore, we propose there is a universal dissipation radius valid in all pair producing pulsars of order $R_{\rm diss}/R_{\rm LC}=\exp\left(\beta^{-1}_{\rm rec}\right)\sim 10^2 - 10^4$, meaning that the stripes should disappear far before reaching the wind termination shock radius in isolated systems like the Crab pulsar where $R_{\rm TS}/R_{\rm LC}\sim 10^9$. 

The wind bulk Lorentz is not as relativistic as previously imagined. After crossing the fast magnetosonic point, the wind speed quickly saturates to $\Gamma_{\infty}\approx\mu^{1/3}_{\rm M}$ and thus does probably not exceed $\Gamma_{\infty}\lesssim 100$ which is closer to gamma-ray burst jet Lorentz factor than the standard Crab pulsar wind model where $\Gamma\sim 10^3-10^6$ \citep{1974MNRAS.167....1R, 1978MNRAS.185..297W, 1984ApJ...283..694K}. On top of the wind acceleration driven by ideal processes, non-thermal particle acceleration proceeds in the striped region driven by reconnection. The particle spectrum is consistent with high-$\sigma$ reconnection simulations, meaning a hard power-law distribution of index $p\sim -1$ between $\gamma_{\rm min}\sim \mu^{1/3}_{\rm M}$ and $\gamma_{\rm max}\sim \mu_{\rm M}$. Scaled up to realistic Crab-like parameters yields the pulsar wind would be composed of ultra-relativistic pairs distributed as $dN/d\gamma\propto\gamma^{-1}$ between $10^2\lesssim\gamma\lesssim10^5$. Injected at the shock front, the wind particles could then naturally explain the mysterious hard radio component in the Crab Nebula \citep{2010A&A...523A...2M}.

\begin{acknowledgements}
We thank the referee, Ioannis Contopoulos, for his careful review of the manuscript. BC warmly thanks Benjamin Crinquand for insightful discussions about this study. This work has been supported by the Programme National des Hautes \'Energies of CNRS/INSU and CNES. We acknowledge PRACE and GENCI (allocation A0070407669) for awarding us access to Joliot-Curie at GENCI@CEA, France.
\end{acknowledgements}

\bibliographystyle{aa}
\bibliography{striped3d}

\end{document}